\documentstyle[12pt,psfig]{article}
\makeatletter
\def\@cite#1#2{{[{#1}]\if@tempswa\typeout
{IJCGA warning: optional citation argument
ignored: `#2'} \fi}}
\newcount\@tempcntc
\def\@citex[#1]#2{\if@filesw\immediate\write\@auxout{\string\citation{#2}}\fi
  \@tempcnta\z@\@tempcntb\m@ne\def\@citea{}\@cite{\@for\@citeb:=#2\do
    {\@ifundefined
       {b@\@citeb}{\@citeo\@tempcntb\m@ne\@citea\def\@citea{,}{\bf ?}\@warning
       {Citation `\@citeb' on page \thepage \space undefined}}%
    {\setbox\z@\hbox{\global\@tempcntc0\csname b@\@citeb\endcsname\relax}%
     \ifnum\@tempcntc=\z@ \@citeo\@tempcntb\m@ne
       \@citea\def\@citea{,}\hbox{\csname b@\@citeb\endcsname}%
     \else
      \advance\@tempcntb\@ne
      \ifnum\@tempcntb=\@tempcntc
      \else\advance\@tempcntb\m@ne\@citeo
      \@tempcnta\@tempcntc\@tempcntb\@tempcntc\fi\fi}}\@citeo}{#1}}
\def\@citeo{\ifnum\@tempcnta>\@tempcntb\else\@citea\def\@citea{,}%
  \ifnum\@tempcnta=\@tempcntb\the\@tempcnta\else
   {\advance\@tempcnta\@ne\ifnum\@tempcnta=\@tempcntb \else \def\@citea{--}\fi
    \advance\@tempcnta\m@ne\the\@tempcnta\@citea\the\@tempcntb}\fi\fi}
\makeatother
\newenvironment{Eqnarray}%
     {\arraycolsep 0.14em\begin{eqnarray}}{\end{eqnarray}}
\def\eq#1{eq.~(\ref{#1})}
\def\Ref#1{Ref.~\cite{#1}}
\def\Refs#1{Refs.~\cite{#1}}
\def\eqs#1#2{eqs.~(\ref{#1})--(\ref{#2})}

\def\eqns#1#2{eqs.~(\ref{#1}) and (\ref{#2})}
\def\simlt{\stackrel{<}{{}_\sim}}
\def\simgt{\stackrel{>}{{}_\sim}}
\def\be{\begin{equation}}
\def\ee{\end{equation}}
\def\bear{\be\begin{array}}
\def\eear{\end{array}\ee}
\def\bea{\begin{Eqnarray}}
\def\eea{\end{Eqnarray}}

\def\mpl{M_{\rm P}}
\def\mx{M_X}
\def\lsim{\mathrel{\raise.3ex\hbox{$<$\kern-.75em\lower1ex\hbox{$\sim$}}}}
\def\gsim{\mathrel{\raise.3ex\hbox{$>$\kern-.75em\lower1ex\hbox{$\sim$}}}}
\def\ifmath#1{\relax\ifmmode #1\else $#1$\fi}
\def\ls#1{\ifmath{_{\lower1.5pt\hbox{$\scriptstyle #1$}}}}
\def\half{\ifmath{{\textstyle{1 \over 2}}}}
\def\threehalf{\ifmath{{\textstyle{3 \over 2}}}}

\def\third{\ifmath{{\textstyle{1 \over 3}}}}
\def\twothirds{\ifmath{{\textstyle{2 \over 3}}}}
\def\tanb{\tan\beta}
\def\hl{h^0}
\def\ha{A^0}
\def\hh{H^0}
\def\hpm{H^\pm}
\def\mha{m_{\ha}}
\def\mhl{m_{\hl}}
\def\mhh{m_{\hh}}

\def\mz{m_Z}
\def\mw{m_W}

\def\mzz{m_Z^2}
\def\mt{m_t}

\def\msusy{M_{\rm S}}
\def\msusyy{M_{\rm S}^2}
\def\mgut{M_{\rm X}}
\def\ie{{\it i.e.}}
\def\eg{{\it e.g.}}

\def\beq{\begin{equation}}
\def\eeq{\end{equation}}
\def\beqa{\begin{Eqnarray}}
\def\eeqa{\end{Eqnarray}}

\def\baselinestretch{1}
\begin{document}
%%%%%%%%%%%%%%%%%%%%%%%%%%%%%%%%%%%%%%%%%%%%%%%%%%%%%%%%%%%%%
\def\IJMPA #1 #2 #3 {{\sl Int.~J.~Mod.~Phys.}~{\bf A#1}\ (19#2) #3}
\def\MPLA #1 #2 #3 {{\sl Mod.~Phys.~Lett.}~{\bf A#1}\ (19#2) #3}
\def\NPB #1 #2 #3 {{\sl Nucl.~Phys.}~{\bf B#1}\ (19#2) #3}
\def\PLB #1 #2 #3 {{\sl Phys.~Lett.}~{\bf B#1}\ (19#2) #3}
\def\PR #1 #2 #3 {{\sl Phys.~Rep.}~{\bf#1}\ (19#2) #3}
\def\PRD #1 #2 #3 {{\sl Phys.~Rev.}~{\bf D#1}\ (19#2) #3}
\def\PTP #1 #2 #3 {{\sl Prog.~Theor.~Phys.}~{\bf #1}\ (19#2) #3}
\def\PRL #1 #2 #3 {{\sl Phys.~Rev.~Lett.}~{\bf#1}\ (19#2) #3}
\def\RMP #1 #2 #3 {{\sl Rev.~Mod.~Phys.}~{\bf#1}\ (19#2) #3}
\def\ZPC #1 #2 #3 {{\sl Z.~Phys.}~{\bf C#1}\ (19#2) #3}
\def\ppnp#1#2#3{{\sl Prog. Part. Nucl. Phys. }{\bf #1} (#2) #3}
\def\npb#1#2#3{{\sl Nucl. Phys. }{\bf B#1} (#2) #3}
\def\jpa#1#2#3{{\sl J. Phys. }{\bf A#1} (#2) #3}
\def\plb#1#2#3{{\sl Phys. Lett. }{\bf B#1} (#2) #3}
\def\prd#1#2#3{{\sl Phys. Rev. }{\bf D#1} (#2) #3}
\def\pR#1#2#3{{\sl Phys. Rev. }{\bf #1} (#2) #3}
\def\prl#1#2#3{{\sl Phys. Rev. Lett. }{\bf #1} (#2) #3}
\def\prc#1#2#3{{\sl Phys. Reports }{\bf #1} (#2) #3}
\def\cpc#1#2#3{{\sl Comp. Phys. Commun. }{\bf #1} (#2) #3}
\def\nim#1#2#3{{\sl Nucl. Inst. Meth. }{\bf #1} (#2) #3}
\def\pr#1#2#3{{\sl Phys. Reports }{\bf #1} (#2) #3}
\def\sovnp#1#2#3{{\sl Sov. J. Nucl. Phys. }{\bf #1} (#2) #3}
\def\jl#1#2#3{{\sl JETP Lett. }{\bf #1} (#2) #3}
\def\jet#1#2#3{{\sl JETP Lett. }{\bf #1} (#2) #3}
\def\zpc#1#2#3{{\sl Z. Phys. }{\bf C#1} (#2) #3}
\def\ptp#1#2#3{{\sl Prog.~Theor.~Phys.~}{\bf #1} (#2) #3}
%%%%%%%%%%%%%%%%%%%%%%%%%%%%%%%%%%%%%%%%%%%%%%%%%%%%%%%%%

%%%%%%%%%%%%%%%%%%%%%%%%%%% subequations.sty %%%%%%%%%%%%%%%%%%%%%%%%
\catcode`@=11
\newtoks\@stequation
\def\subequations{\refstepcounter{equation}%
\edef\@savedequation{\the\c@equation}%
  \@stequation=\expandafter{\theequation}%   %only want \theequation
  \edef\@savedtheequation{\the\@stequation}% % expanded once
  \edef\oldtheequation{\theequation}%
  \setcounter{equation}{0}%
  \def\theequation{\oldtheequation\alph{equation}}}
\def\endsubequations{\setcounter{equation}{\@savedequation}%
  \@stequation=\expandafter{\@savedtheequation}%
  \edef\theequation{\the\@stequation}\global\@ignoretrue

\noindent}
\catcode`@=12
%%%%%%%%%%%%%%%%%%%%%%%%%%%%%%%%%%%%%%%%%%%%%%%%%%%%%%%%%%%%%%%%%%%%%
\begin{titlepage}

%\title{{\bf The MSSM infrared fixed point scenario, the Higgs mass
%and LEP--2}
\title{{\bf The Higgs mass in the MSSM infrared fixed point scenario}%
\thanks{Research supported in part by: the CICYT, under
contract AEN95-0195 (JAC); the European Union,
under contract CHRX-CT92-0004 (JAC); and the U.S. Department of Energy,
grant DE-FG03-92ER40689.}
}
\author{ {\bf J.A. Casas${}^{ {\footnotesize\P}}$},
{\bf J.R. Espinosa${}^{ {\footnotesize\dag}}$ }
and {\bf H.E. Haber${}^{ {\footnotesize\S}}$}\\
\hspace{3cm}\\
${}^{\footnotesize\P}$ {\small Instituto de
Estructura de la Materia, CSIC}\\
{\small Serrano 123, 28006 Madrid, Spain}\\
{\small casas@cc.csic.es}\\
${}^{\footnotesize\dag}$ {\small CERN, TH Division} \\
{\small CH-1211 Geneva 23, Switzerland}\\
{\small espinosa@mail.cern.ch}\\
${}^{\footnotesize\S}$ {\small Santa Cruz Institute for Particle Physics}\\
{\small University of California, Santa Cruz, CA 95064, USA}\\
{\small haber@scipp.ucsc.edu}}
%\vspace{-0.3cm}\\
\date{}
\maketitle
\def\baselinestretch{1.15}
\begin{abstract}
\noindent
In the infrared fixed point (IFP) scenario of the minimal supersymmetric
model (MSSM), the top-quark mass and other physical quantities of the
low-energy theory are insensitive to the values of the parameters
of the theory at some high energy scale.
%resulting in a highly predictive low-energy scenario.
In this framework we evaluate the light CP-even Higgs mass, $\mhl$,
taking into account some important effects that had not
been previously considered.
%The most notable of these
%is the supersymmetric correction to the relation
%between the running and the physical top-quark masses, which
%lowers the value of $\tan\beta$.  As a result, the predicted value of
%$\mhl$ is significantly lower than in previous evaluations.
In particular, the supersymmetric correction to the relation
between the running and the physical top-quark masses
lowers the value of $\tan\beta$, thereby implying a lower predicted
value of $\mhl$.
Assuming a supersymmetric threshold of $M_S\leq 1$ TeV and $M_t=175$~GeV,
we find an upper bound of
$\mhl\le 97\pm 2$~GeV; the most plausible value of $\mhl$ lies somewhat
below the upper bound.  This places the Higgs boson
in the IFP scenario well within the reach of the LEP-2 Higgs search.

\end{abstract}

\thispagestyle{empty}

\leftline{}
\leftline{CERN-TH/98-12}
\leftline{January 1998}
\leftline{}

\vskip-23.9cm
\rightline{}
\rightline{IEM-FT-167/98}
\rightline{CERN-TH/98-12}
\rightline{SCIPP-98-01}
\rightline{hep-ph/9801365}
\vskip3in

\end{titlepage}
%%%%%%%%%%%%%%%%%%%%%%%%%%%%%%%%%%%%%%%%%%%%%%%%%%%%%%%%%%%%%%%%%%%
%%%%%%%%%%%%%%%%%%%%%%%%%%%%%%%%%%%%%%%%%%%%%%%%%%%%%%%%%%%%%%%%%%%
\newpage
\setcounter{page}{1}

\section{Introduction}

Models of low-energy supersymmetry can add many new parameters to the
Standard Model.  The minimal supersymmetric extension of the Standard
Model (MSSM) is minimal only in its choice of particle content.  The
number of free parameters of the model is quite large unless
additional theoretical assumptions are imposed.  The parameter freedom
of the MSSM is due mostly to soft supersymmetry-breaking
parameters, whose theoretical origins are unknown.
It is common practice to
treat the parameters of the MSSM as running parameters and impose a
particular structure on the soft supersymmetry breaking terms at
a common high energy scale [such as the Planck
scale ($\mpl$) or grand unification (GUT) scale ($\mx$)].
Using the renormalization group equations (RGEs), one can then
derive the values of the low-energy MSSM parameters.

A particularly attractive framework, which we will adopt in this
paper, consists of assuming universality of soft breaking
parameters at the high-energy unifying scale. Universality
is a desirable property not only to reduce the number of
independent model parameters, but also to avoid unsuppressed
flavor changing neutral currents (FCNCs) \cite{fcnc}.
Universality of scalar and gaugino mass parameters in the high energy
theory is an automatic consequence of
the {\it minimal} supergravity (SUGRA) framework \cite{Nilles}
and approximately holds
in several string-derived SUGRA scenarios \cite{strings}.

The resulting low-energy supersymmetric theory that emerges depends on
five supersymmetric model parameters: a common scalar mass $m$, a
common gaugino mass $M$, a common flavor-diagonal
trilinear scalar coupling $A$,
a supersymmetric Higgs mass parameter $\mu$, and an
off-diagonal Higgs squared-mass parameter $m_{12}^2$ (often called
$B\mu$).  These parameters are high-energy scale parameters (defined
at either $\mgut$ or $\mpl$) and serve as initial conditions for the
RGEs.  Electroweak symmetry breaking in the low-energy theory is
radiatively generated when one of the Higgs squared-masses is driven
negative by renormalization group (RG) running.  Then, by imposing
the minimum conditions for the Higgs potential, one can eliminate
$\mu^2$ and $m_{12}^2$ in favor of the electroweak symmetry-breaking
scale, $v^2\equiv v_1^2+v_2^2\simeq (174~{\rm GeV})^2$, and
$\tan\beta\equiv v_2/v_1$, where $v_1$ and $v_2$ are the Higgs vacuum
expectation values.
% (this will be explained in more detail in section 2).
The sign of $\mu$ is not fixed in this procedure and remains a free
parameter.

Clearly, the previously described SUGRA theory
is a highly constrained version of the MSSM.  Nevertheless,
there can be additional interesting constraints. In particular,
certain low-energy MSSM parameters are sometimes very insensitive to the
initial high energy values of
the SUGRA parameters.  Such a possibility is very exciting, since it
offers a potential for understanding the physical value of some
low-energy parameters without a detailed knowledge of the physics at
high energies.

The classic example of the scenario just described is the quasi-infrared
fixed point (IFP) prediction for the top-quark Yukawa coupling
\cite{hill,madrid,BBOP,cw,cw2,ross,ifprev,PaulNir,Nir,chapo,bb,Abel}%
\footnote{The quasi-infrared fixed point differs from the infrared
fixed point of Pendleton and Ross (PR)\cite{pendleton}.  The PR fixed
point is an infrared stable fixed point that is reached at a scale
$Q$ for sufficiently large $\mx/Q$.  However, in practice $\mx/\mz$ is
not large enough, so the PR fixed point solution does not govern the
low-energy value of the top-quark Yukawa coupling.  On the other hand,
it follows from \eqs{Yt}{IFP} that the top-quark Yukawa coupling
is driven to the quasi-infrared fixed point as long as $Y_t(0)F(t_Z)\gg
1/6$, where $t_Z\equiv\ln(\mx^2/\mz^2)$.}.
As is well known \cite{madrid}, the one-loop RGE of the top-quark
Yukawa
coupling,
$Y_t\!\equiv\! h_t^2/(4\pi)^2$, can be integrated analytically for
moderate values of $\tan\beta\sim{\cal O}(1)$:
\be
\label{Yt}
Y_t(t)=\frac{Y_t(0)E(t)}{1+6Y_t(0)F(t)},
\ee
with
\be
\label{EF}
E(t)= (1+\beta_3t)^{16/3b_3}(1+\beta_2t)^{3/b_2}
(1+\beta_1t)^{13/9b_1}\;,\;\;F(t)=\int_0^tE(t')dt'\,.
\ee
In \eq{EF},
$\beta_i\equiv\alpha_i(0)b_i/4\pi$ are the one-loop beta functions of
the gauge couplings $\alpha_i(t)$, with $(b_1,b_2,b_3)=(11,1,-3)$,
and $t=\ln (M_X^2/Q^2)$, where $Q$ is
the renormalization scale.
This one-loop behavior leads to the existence of the quasi-infrared
fixed point. Namely, for $Y_t(0)\rightarrow\infty$,
\be
\label{IFP}
Y_t(t)\rightarrow Y_f(t)\equiv\frac{E(t)}{6F(t)}.
\ee

Numerically, one finds that $Y_t$ at the electroweak scale differs
negligibly from $Y_f$ for a wide range of $Y_t(0)\gsim 0.01$, so in
this sense the low-energy value of $Y_t$ is indeed insensitive to its
high-energy value $Y_t(0)$.  The value of the top-quark mass depends
both on the low-energy values of $Y_t$ and $\tan\beta$, so at this
stage we do not have a prediction for the top-quark mass.
Nevertheless, the parameter freedom has been reduced, since given the
top-quark mass, $\tan\beta$ is now predicted.
Actually, $\tan\beta$ typically turns out to be near 1,
in which case the previous derivation is fully justified\footnote{
If one solves the complete set of RGEs for the top
and bottom-quark Yukawa couplings, one finds another IFP solution
with $\tan\beta\sim m_t/m_b$. In this paper, we will not
address this large $\tan\beta$ scenario since, in the minimal SUGRA
approach described above, it requires a precise (unnatural)
fine-tuning of high-energy parameters in order to ensure
the correct radiative electroweak symmetry breaking \cite{unnatural}.}.

In this paper, we focus on the prediction of the light CP-even
Higgs mass ($\mhl$) in the
IFP scenario as a function of the minimal SUGRA parameters.  We improve
on
previous work in the literature by taking into account a number of effects
not fully considered before.  These include:
(i) corrections to $\tan\beta$ due to supersymmetric
thresholds; (ii) evolution of $\tan\beta$ from the electroweak scale to
the supersymmetry-breaking scale; and (iii) a precise evaluation of
radiative
electroweak breaking and of the top-squark (stop) mixing parameter. All
these effects have
a significant impact on the value of $\mhl$. In addition, we have
computed $\mhl$ using the most refined methods available, including
subdominant radiative corrections and contributions from
stop non-degeneracy. This
substantially reduces the theoretical uncertainty of our results with
respect to previous literature. Our final result on the upper bound on
the Higgs mass
%is about 10 GeV smaller than in previous evaluations.
%as we will see.
%This is a substantial decrease,
%which
has important implications for the LEP-2 Higgs search.

In Section~2, we discuss the IFP scenario and the calculation of
$\tan\beta$, as well as the stop mixing parameter, including all
the new effects mentioned above.  We
address a number of effects not previously considered,
which can significantly affect the predicted value of
$\tan\beta$ and the Higgs mass. In Section~3, we review
the dependence of the Higgs mass on the
supersymmetric parameters.  In Section~4, we explore the consequences
of the IFP scenario for the predicted value of the Higgs
mass, giving full numerical results and comparing to the previous
literature. Conclusions are presented in Section~5.

\section{The IFP scenario revisited}

In Section~1, we reviewed the quasi-infrared fixed point (IFP) scenario
in which the low-energy value of the top-quark Higgs Yukawa coupling
is driven to a quasi-infrared fixed point value, $Y_f$.  Formally,
this limit is derived by taking $Y_t(0)\to\infty$.  This is not
theoretically consistent as it stands, since the derivation given
above was based on a one-loop RGE, while large values of $Y_t(0)$
clearly lie outside the
perturbative regime.
However, it has been shown \cite{Jones} that the domain of attraction of
the quasi-IFP is large and accurately represented by the one-loop
approximation. In particular, $Y_t(0)$ rapidly approaches $Y_f$,
even for values of $Y_t(0)$ still in the perturbative region.
This allows one to consider the IFP limit as a meaningful physical
possibility. For example,
starting with $Y_t(0)=0.1$ the one-loop value of
$Y_t(t)$ at the weak scale differs from $Y_f$ by $0.27\%$.
In this paper, we employ two-loop RGEs for the evolution of the
gauge and Yukawa couplings.  For definiteness, we choose
$Y_t(0)=0.1$, although the results are insensitive to this choice,
as argued above.

Another subtlety concerning the precise definition of the IFP
scenario is the choice of the unification scale $M_X$ and of
$\alpha_i(0)$.  Here,
we follow the approach of \Ref{mssmpc}.  First, we
take the experimental values of $\alpha_i(Q\!=\!\mz)$ as input
parameters and evaluate the corresponding supersymmetric
$\overline{\rm{DR}}$ values, $\hat \alpha_i(\mz)$, taking into account all
the supersymmetric threshold corrections\footnote{Of course, these
threshold corrections depend on the values of supersymmetric masses
and thus on the remaining independent parameters of the model.}
[the $\hat \alpha_i(\mz)$ do not have a direct
physical meaning; see \Ref{mssmpc} for more details]. Then, the
two-loop running of
$\hat\alpha_1(t)$, $\hat\alpha_2(t)$ to high scales defines a
unification scale $M_X$ and a ``unified" coupling constant
$\hat\alpha(0)$. Finally, the running of $\hat\alpha_3$ from $\mz$ to
$M_X$ gives the value of $\hat \alpha_3(0)$. In general, the latter
does not coincide (even within the error bars) with $\hat \alpha(0)$,
although the difference is small and can be attributed to, for instance,
threshold corrections either from a GUT or stringy origin.

The IFP scenario defined in the context of the SUGRA approach
depends on additional parameters $m$, $M$, $A$, $\tan\beta$
and sign($\mu$) as described in Section~1.  However,
the subset of independent parameters is substantially smaller. In the
IFP scenario, the low-energy value of $A_t$ (the trilinear scalar
coupling of the Higgs boson and stops) is also
driven to an infrared quasi-fixed point \cite{cw}. At the one-loop level
\be
\label{At}
A_t(t)\rightarrow M
\left[\frac{1}{4\pi}\left(\frac{16}{3}\alpha_3(0)h_3 +
3\alpha_2(0)h_2 + \frac{13}{9}\alpha_1(0)h_3\right)
- t\frac{E(t)}{F(t)} +1 \right],
\ee
where $h_i(t)=t/(1+\beta_it)$. Therefore the value of $A_0$ in the
IFP limit is irrelevant. Although this is not true for the remaining
trilinear couplings $A_b, A_\tau$, etc., the latter $A$-parameters
have a negligible effect in the determination of the Higgs
mass, which is the main goal of this paper. The value of
$\tan\beta$, evaluated at the scale $Q=M_t$
(where $M_t$ is the physical top-quark mass), is determined by using
\be
\label{H2}
v_2(M_t) = \frac{m_t(M_t)}{4\pi\sqrt{Y_t(M_t)}}\,,
\ee
and the approximate $\overline{\rm DR}$ relation
\cite{mssmpc}
\be
\label{v2}
v(\mz)\simeq
\left[175.8+0.32\ln\left({m^2+4M^2\over \mz^2}\right)\right]\;{\rm GeV}.
\ee

The distinction between the physical top-quark mass $M_t$ and the
running top-quark mass $m_t(M_t)$ should not be ignored.  Explicitly,
the physical top-quark mass is given by
\be
\label{Mt}
M_t=m_t(M_t)\left[1+\frac{\Delta m_t}{m_t}\right],
\ee
where the one-loop correction $\Delta m_t$ receives two important
contributions:
the well-known QCD gluon correction\footnote{The factor 5 in
\eq{DmtQCD}\ in the $\overline{{\rm DR}}$ scheme should be compared
with 4 in the $\overline{{\rm MS}}$ scheme \cite{mssmpc}.}
\be
\label{DmtQCD}
\left(\frac{\Delta m_t}{m_t}\right)_{\rm QCD}=\frac{5g_3^2}{12\pi^2}\,,
%\left[5+3\log\frac{Q^2}{m_t^2}\right]
\ee
and the stop/gluino correction \cite{mssmpc,thrqcd}
\bea
\label{DmtSUSY}
\left(\frac{\Delta m_t}{m_t}\right)_{\rm SUSY}=
&&\left.-\frac{g_3^2}{12\pi^2}
\right\{B_1(m_t,M_{\tilde g}, m_{\tilde t_1})
+B_1(m_t,M_{\tilde g}, m_{\tilde t_2})
\nonumber\\
&&-\left.\sin(2\theta_t)\frac{M_{\tilde g}}{m_t}
\left[B_0(m_t,M_{\tilde g}, m_{\tilde t_1})
-B_0(m_t,M_{\tilde g}, m_{\tilde t_2}) \right]
\right\}\,,
\eea
where $\theta_t$ is the stop mixing angle, $m_{\tilde t_1}>
m_{\tilde t_2}$, and
\be
B_n(p\,;m_1,m_2)\equiv -\int_0^1 dx\, x^n\,\ln\left[{(1-x)m_1^2+xm_2^2
-x(1-x)p^2\over m_t^2}\right]\,.
\ee
%the definition of the
%$B_0, B_1$ loop-functions can be found in \Ref{mssmpc}.
Note that the Standard Model two-loop QCD correction
\cite{2QCD} and the electroweak correction \cite{hb}
are each of order 1\% and almost cancel one another.  While the
one-loop gluon correction [\eq{DmtQCD}] yields a 6\% relative top-quark
mass
shift, the supersymmetric correction in our scenario is of the
same sign and can be as
large as the gluon correction for
$M\gsim 500$ GeV. The stop/gluino
correction (which increases with the supersymmetric masses) is a
consequence of working in the effective supersymmetric theory without
decoupling the supersymmetric particles, as is usually done in the IFP
literature when considering the running of $Y_t$.
(In practice, this is the most convenient way to perform the analysis;
for an alternative approach, see \Ref{fpdec}.)  However, the
correction given by \eq{DmtSUSY}\ has never been included in the
published analyses of the IFP scenario.
This correction has the noteworthy effect of
reducing the ratio $m_t(M_t)/M_t$, and consequently lowering the IFP
value of $\tan\beta$.  As a result, the predicted value for the mass of
the light CP-even Higgs boson is significantly reduced, as shown in
Section~4.

Let us now turn to the $\mu$-parameter.
We noted in Section~1 that $\mu$ can be determined (up
to a sign) by imposing the condition of electroweak
symmetry breaking and fixing the $Z$ mass to its physical value.
More precisely, from the minimization of the
renormalization-group-improved tree-level Higgs potential, we obtain
\be
\label{mu}
\mu^2+\half\mzz=\frac{1}{\tan^2\beta-1}\left(
m_{H_1}^2 - \tan^2\beta\ m_{H_2}^2 \right),
\ee
where $m_{H_1}^2,m_{H_2}^2$ are the low-energy values of the soft
squared-masses of the $H_1,H_2$
Higgs fields (subject to the condition $m_{H_1}^2\!=\!m_{H_2}^2\!=\!m^2$
at $Q\!=\!M_X$). It is important to note that the result given in
\eq{mu} is only
accurate enough if the tree-level potential is evaluated at a scale
where the radiative corrections are minimized. This essentially
happens for a scale of order the stop masses \cite{Gambe,Bea}. From now on
we will take that scale, $M_S$, as the average of the stop squared-mass
eigenvalues
\be
\label{Ms}
M_S^2\equiv\half\left( m_{\tilde{t}_1}^2 + m_{\tilde{t}_2}^2\right)\,.
\ee
Consequently, all the quantities appearing in \eq{mu}\ (including
$\mu$ and $\mz$) are to be taken at $Q\!=\!M_S$.\footnote{Even
including the one-loop radiative corrections $\Delta V_1$
to the tree-level potential $V_0$, and using \eq{mu}\
accordingly modified, is not in general a precise procedure since
$V_0+\Delta V_1$ at $Q=\mz$ still yields inaccurate results if $M_S^2\gg
\mzz$, as it is normally the case \cite{Bea} (see the
comments at the end of the appendix).} {}From \eq{H2}, \eq{v2}\ {\em plus}
the renormalization group
evolution of $H_1$, $H_2$
with their anomalous dimensions, we can determine the value of $\tan
\beta$ at any scale using the corresponding RGE for $\tan\beta$:
\be
\label{tbetaRGE}
\frac{d\tan\beta}{dt}\simeq \threehalf Y_t\tan\beta\,.
\ee
This result can be employed to determine the value of $\tan\beta$ at
$M_S$.  The running of $\tan\beta$ has been
ignored in the IFP literature and produces significant
corrections in the final results.

%Again, this is something
%that has
%been frequently disregarded in the IFP literature.

{}From \eqs{At}{v2}\ and \eq{mu}\ it follows
that the only relevant independent parameters for predicting the light
CP-even Higgs mass $\mhl$ in the IFP scenario are $m$ and $M$.
%(this set could be somewhat
%enhanced by allowing non--universal boundary conditions
%for $m$).
These can be traded in for $M_S$ and
$x\equiv
M/m$. Notice that in either case sign$(\mu)$ may be absorbed, by a
redefinition of fields, into the sign of $M$ (or equivalently, the
sign of $x$). Besides the
simplicity of this scenario, the fact that all the relevant low-energy
quantities can be expressed in terms of $M_S$ and $x$ has important
consequences for the prediction of $\mhl$. In
particular, the mass splitting between stops and the effective
mixing\footnote{The convention for the sign of $\mu$ in \eq{Xt}\ is
opposite to the one employed in \Ref{pdg}.}
\be
\label{Xt}
X_t\equiv A_t+\mu\cot\beta\,,
\ee
which play an important role in the computation
of $\mhl$ (see Section~3) are no longer independent parameters,
but are calculable quantities in terms of $M_S$ and $x$. Since they
cannot be simultaneously ``tuned" to
the values that maximize $\mhl$, this produces
an effective lowering of the upper bound on $\mhl$.
These issues will be carefully analyzed
in the next two sections.

There is yet another source of constraints
on the theory, namely the desirable absence of dangerous charge and
color breaking (CCB) minima \cite{CCB,CLM} or unbounded from below (UFB)
directions \cite{CLM}
in the scalar potential. CCB and UFB constraints have been
recently analyzed for
the IFP scenario \cite{CCBIFP}. Since all the physics in which we are
interested depends on just two parameters, $M$ and $m$ (or equivalently
$M_S$ and $x$), we must focus on the CCB and UFB constraints involving
these quantities. This means, in particular, that
the CCB constraints involving the trilinear scalar couplings other
than the top one, \ie $\,$ $A_u, A_d, A_s,$ etc., have no relevance to us
since their low-energy values may be tuned at will by varying
the initial high-energy parameter $A_0$. This is not the case
for the low-energy top trilinear scalar coupling $A_t$, which
in our scenario
is driven to an infrared fixed point given
by \eq{At}\ [more generally, by \eq{At2}],
namely $A_t\simeq -1.2 M$. This value, however, is well inside
the region allowed
by the CCB bounds \cite{CCBIFP}. On the other hand, UFB bounds strongly
restrict the $x\equiv M/m$ parameter \cite{CCBIFP} in the IFP
scenario, namely the absence of UFB directions requires $|x|\leq 1$.
In any case, the results presented in Section~4 imply that
for $x>1$ the value of $\mhl$ hardly changes as a function of $x$
(\ie $\,$it already reaches its large-$x$ asymptotic limit at $x=1$).
Thus, in
practice the CCB and UFB constraints do not restrict the bounds on $\mhl$
in the IFP scenario.

\section{The MSSM Higgs mass}

The Higgs sector of the MSSM consists of five physical states: two
neutral CP-even scalars $\hl$ and $\hh$ (with $\mhl\leq\mhh$), one
neutral CP-odd scalar $\ha$, and a charged Higgs pair $\hpm$.
The quadratic terms of the Higgs potential consists of two
diagonal squared-mass terms: $m_i^2\equiv m_{H_i}^2 +|\mu|^2$,
and one
off-diagonal squared-mass term: $m_{12}^2$.  When the minimum condition
is imposed, the diagonal squared-mass terms are traded in for the vacuum
expectation values $v_1$ and $v_2$.  Thus, the tree-level Higgs sector
depends on only two new parameters: $\tan\beta$ and $m_{12}^2$.  It is
convenient to replace $m_{12}^2$ with the physical parameter $\mha$.
Then, all other Higgs masses and couplings can be expressed at
tree level in terms of $\tan\beta$ and $\mha$.

The prediction for the mass of the lightest CP-even neutral Higgs boson
is of particular interest to the LEP Higgs search, since this Higgs
scalar would be discovered first if it lies within the reach of the
LEP-2 collider.  In particular,
the MSSM predicts that at tree level, $\mhl\leq \mz|\cos 2\beta|\leq
\mz$. When radiative corrections are included, the
light Higgs mass upper bound may be significantly increased above the
tree level prediction.  This has profound effects on the LEP Higgs
search.  LEP-2 running at its maximum energy ($\sqrt{s}\simeq 200$~GeV)
and luminosity is expected
to be sensitive to Higgs masses up to about 105 GeV \cite{yellowbook}.
Thus, the
possibility of large radiative corrections to $\mhl$ implies that LEP
cannot be sensitive to the full MSSM Higgs sector parameter space.

The mass of $\hl$
can be calculated in terms of the two parameters of
the Higgs sector mentioned above ($m_{A^0}$ and $\tan\beta$) and other
MSSM
soft-supersymmetry-breaking parameters that affect the Higgs mass
through virtual loops
\cite{hhprl,veff,hempfhoang,completeoneloop,rge,2loopquiros,ceqr,carena,%
carena2,hhh}.
The largest contribution to the one-loop radiative
corrections is enhanced by a factor of $m_t^4$ and grows
logarithmically with the stop mass.  Thus, higher-order
radiative corrections can be non-negligible for large stop
masses, in which case
the large logarithms must be resummed using
renormalization group techniques
\cite{rge,2loopquiros,ceqr,carena,carena2,hhh}.

For our numerical work, we will follow the simple analytic procedure
for accurately approximating $\mhl$ described in \Ref{hhh}, where
further details can be found. Similar results are obtained by using the
alternative approximation of \Refs{carena,carena2}. These analytic
formulae
incorporate both the leading one-loop and two-loop effects and the
RG improvement. Also included are the leading effects at one loop of the
supersymmetric thresholds (the most important effects of this type are
squark mixing effects in the third generation).

In the limit $\mha\gg \mz$, which holds in the IFP scenario, only $\hl$
remains light (with couplings nearly identical to those of the Standard
Model Higgs boson), and its squared-mass including RG improvement is
given by a formula of the form\footnote{Corrections associated
with sbottom virtual loops are small if $\tan\beta$ is small, and so they
are not shown explicitly in \eq{mhhhh}, although they were
included in our numerical analysis.} \cite{hhh}
\be
\label{mhhhh}
\mhl^2\simeq
({\mhl^2})_{\rm 1LL}\left[m_t(\mu_t)\right]+
(\Delta\mhl^2)_{\rm mix}\left[m_t(\mu_{\tilde t})\right]\,,
\ee
where
\be
\mu_t\equiv\sqrt{m_t\msusy}\,,\qquad\mu_{\tilde t}\equiv M_S\,.
\ee
In particular, the numerically integrated RG-improved
CP-even Higgs mass is well approximated by replacing all
occurrences of $m_t$ in
$({\mhl^2})_{\rm 1LL}$ and  $(\Delta\mhl^2)_{\rm mix}$ by
the corresponding running masses evaluated at the
judicious choice of scales indicated above.

The first term in \eq{mhhhh}\ is the one-loop leading logarithmic
contribution to the squared mass, given by
\be
\label{mh1ll}
({\mhl^2})_{\rm 1LL}\simeq \mzz\cos^22\beta+
\frac{g^2 N_c \mt^4(\mu_t)}{8\pi^2\mw^2}
\ln\left(\frac{m_{\tilde{t}_1} m_{\tilde{t}_2}}{m_t^2}\right)\,,
\ee
where $\tan\beta$ is evaluated at $\mz$ and $N_c=3$.
Subdominant terms not written in \eq{mh1ll}\ can also be important
for a
precise determination of $\mhl$. They can be found in
\Ref{hhh} and were included in our numerical analysis.
The second term in \eq{mhhhh}\ adds the important effects of
stop mixing; it takes the form (again we display here only the dominant
terms)
\be
\label{deltamix}
(\Delta\mhl^2)_{\rm mix}
\simeq\frac{g^2N_c}{16\pi^2m_W^2}m_t^4(\mu_{\tilde t})\left\{X_t^2
\left[2h( m_{\tilde{t}_1}^2 , m_{\tilde{t}_2}^2)
+X_t^2g( m_{\tilde{t}_1}^2 , m_{\tilde{t}_2}^2)\,
\right]\right\},
\ee
where $X_t$ is given by \eq{Xt}, and
\bear{cl}
h(a,b)&={\displaystyle {1\over a-b}\ln{a\over b}}\,,\nonumber\\[.1in]
g(a,b)&={\displaystyle {1\over (a-b)^2}\left(
2-{a+b\over a-b}\ln{a\over b}
\right)}\,.
\eear
Using these results,
the full (numerically integrated) RG-improved value of
$\mhl$ is reproduced to within an accuracy of about 2~GeV (assuming that
supersymmetric particle masses lie below 2 TeV).

For $|m_{\tilde{t}_1}^2-m_{\tilde{t}_2}^2|\ll M_S^2$, we may approximate
$g(a,a)\simeq -1/6a^2$ and $h(a,a)\simeq 1/a$.  Then
\eq{deltamix}\ simplifies and takes the form
\begin{eqnarray} \label{deltamhs}
(\Delta\mhl^2)_{\rm mix}\!&=&\!{g^2 N_c\over 16\pi^2\mw^2}
m_t^4 \left\{{2 X_t^2\over \msusyy}\left(\!1-{X_t^2\over
12\msusyy}\right)\right\}\,.
\end{eqnarray}
{}From \eq{deltamhs}\ it is easy to see that the maximal value of
$(\Delta\mhl^2)_{\rm mix}$, and thus
$\mhl$, is achieved for $|X_t| = \sqrt{6}\msusy$, which
is sometimes called ``maximal mixing''.
%\footnote{If the stop mass
%splitting is primarily due to $X_t$, then the approximation which leads
%to \eq{deltamhs} is valid if $m_t |X_t|/M_S^2\ll 1$.  For example, for
%$M_S=1$~TeV, values of |X_t|/M_S\lsim 3$ (which includes the point of
%maximal mixing) should yield an acceptable approximation based on
%\eq{deltamhs}.}.
For this value of $|X_t|$,
the quantity in curly brackets in \eq{deltamhs}\
is equal to 6. For larger values of $|X_t|$ this correction decreases,
eventually turning negative. In the IFP scenario the approximation of
nearly degenerate stops is not always applicable
(particularly for small values of $|x|$, as shown in fig.~1), and one
must include the stop mixing corrections in its full form
[\eq{deltamix}].  In the latter case,
$(\Delta\mhl^2)_{\rm mix}$ does not follow the simple
behaviour discussed for the approximately mass-degenerate case; for
example, values larger
than 6 for the term in curly brackets in \eq{deltamix}\ can result.

 %%%%%%%%%%%%%%%%%%%%%%%%figure%%%%%%%%%%%%%%%%%%%%%%%%
\begin{figure}[hbt]
%%\psdraft
\centerline{
\psfig{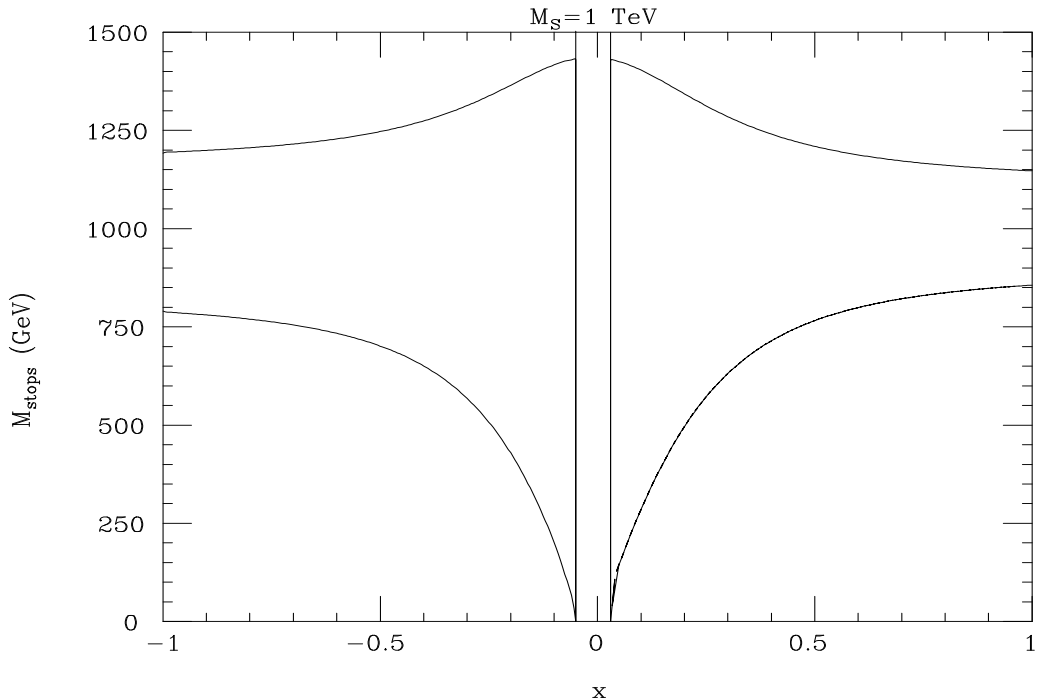}}
\caption{\footnotesize Stop-mass eigenvalues $m_{\tilde{t}_1}$
(upper curves), $m_{\tilde{t}_2}$ (lower curves),
as a function of $x=M/m$ in the IFP scenario for
$M_S=1$~TeV and $M_t=175$~GeV. 
The indicated area around $x=0$ has one very light stop
(or a negative value of $m^2_{\tilde{t}_2}$)
and is thus excluded experimentally.}
\end{figure}
%%%%%%%%%%%%%%%%%%%%%%%%figure%%%%%%%%%%%%%%%%%%%%%%%%

As an example, one finds the following mass bounds for $\hl$, assuming
$M_t=175$~GeV and $M_{S}\lsim 1$~TeV:
$\mhl\lsim 112$~GeV if stop mixing is negligible, while
$\mhl\lsim 125$~GeV if stop mixing is ``maximal''.
In both cases the upper bound
corresponds to large $\tan\beta$.
When the IFP scenario is imposed, the parameter restrictions
examined in Section~2 (\eg$\,$ both $\tan\beta$ and $A_t$ are
driven to fixed-point values) imply that the Higgs mass
upper limits quoted above are not saturated.
Consequently the predicted value of $\mhl$ decreases substantially.
In Section~4, we shall explore in detail the predictions for $\mhl$
in the IFP scenario as a
function of the remaining free parameters.

\section{Results}

For the sake of definiteness and to facilitate the comparison with
previous results in the literature, we first present
detailed results for
$M_S=1$ TeV. Subsequently, we will allow $M_S$ to
vary. It is then illustrative to start by
showing the dependence of several relevant quantities
as a function of the only
remaining parameter, $x \equiv M/m$. In all the cases we will
vary $x$ over the range $[-1,1]$, since for
$|x|\ge 1$ all the relevant quantities enter an asymptotic
regime, as will be apparent from the figures. In addition, as
explained at the end of Section~2, the values $|x|\simgt 1$
are in conflict with CCB and UFB bounds.

In fig.~1 we plot
the two stop mass eigenvalues $m_{\tilde t_1},m_{\tilde t_2}$ as a
function of $x$.
We note that for $-0.07\simlt x \simlt 0.03$ the mass of the
lightest stop is lower than the present
experimental bounds \cite{pdg2}. Thus, this region is excluded, as
indicated in all figures shown in this section.  We also observe that
\eq{deltamhs} is no longer a good approximation for
$(\Delta\mhl^2)_{\rm mix}$ when $|x|\lsim 0.4$,
and one must use \eq{deltamix}, as noted at the end of Section~3.

%%%%%%%%%%%%%%%%%%%%%%%%figure%%%%%%%%%%%%%%%%%%%%%%%%
\begin{figure}[hbt]
%%\psdraft
\centerline{
\psfig{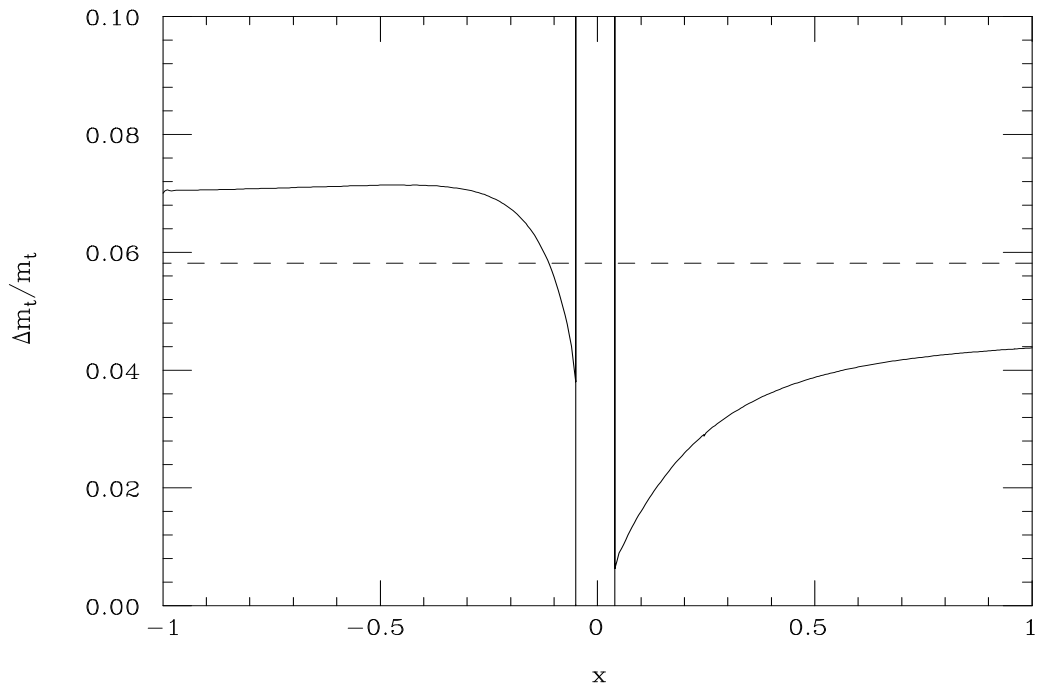}}
\caption{\footnotesize
Stop/gluino threshold contribution [\eq{DmtSUSY}] to the
relationship between the physical (pole) top-quark mass $M_t$ and the
$\overline{\rm DR}$ running mass $m_t\equiv m_t(M_t)$, as a function of
$x\equiv M/m$ in the IFP scenario with $M_S=1$~TeV 
and $M_t=175$~GeV (solid line).  For comparison, the gluon
contribution [\eq{DmtQCD}] is also exhibited (dashed line).}
\end{figure}
%%%%%%%%%%%%%%%%%%%%%%%%figure%%%%%%%%%%%%%%%%%%%%%%%%

Figure~2 shows the supersymmetric
correction (due to stop/gluino loops)
to the top-quark mass $(\Delta m_t/m_t)_{\rm SUSY}$ as a function of $x$.
Note that this correction is in general
quite important. For comparison, we have also plotted the usual
QCD correction, $(\Delta m_t/m_t)_{\rm QCD}$
(constant dashed line).  Although the
supersymmetric correction does not always have a definite sign
in general models (as noted in \Ref{Nir}), this correction is always
of the same sign as the QCD correction in the IFP scenario
considered in this paper.  This feature is a result of
the constraints imposed on the stop and gluino masses. Moreover, the
larger the positive value of $\Delta m_t$,
the lower the value of $\tan\beta$.  This can be seen from the dashed
line in fig.~3, which shows the behaviour of $\tanb$ as a function of
$x$.

%%%%%%%%%%%%%%%%%%%%%%%%figure%%%%%%%%%%%%%%%%%%%%%%%%
\begin{figure}[hbt]
%%\psdraft
\centerline{
\psfig{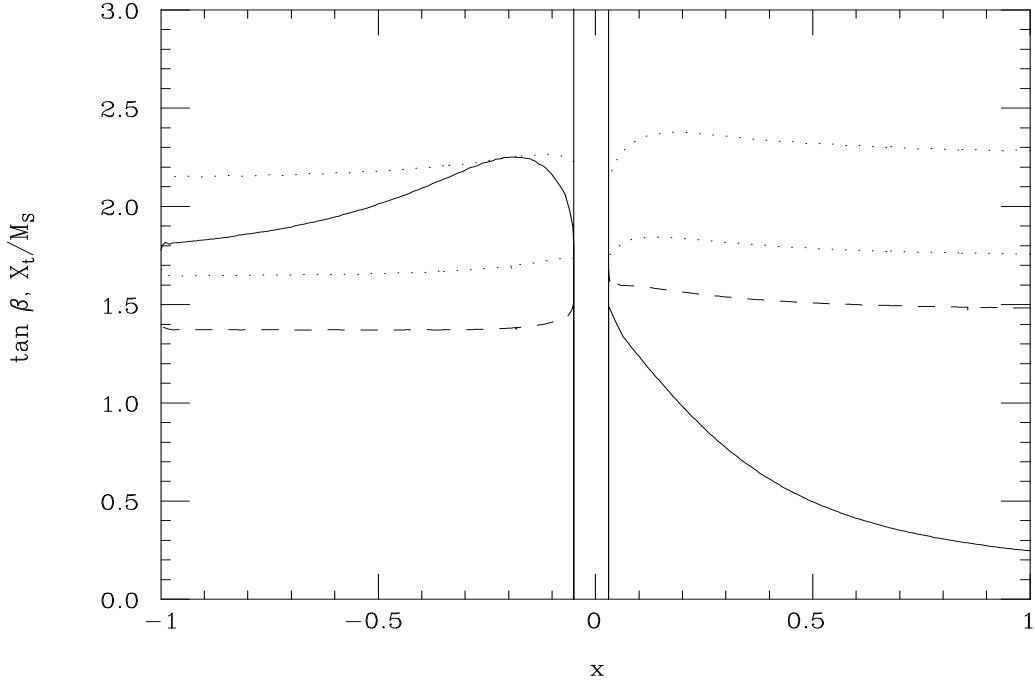}}
\caption{\footnotesize 
Predicted value of the stop mixing parameter $X_t/M_S$ (solid line) and
$\tan\beta$ (dashed line) in the IFP scenario as a function of
$x\equiv M/m$ with $M_S=1$~TeV and $M_t=175$~GeV.
Both parameters are essential in
the determination of $\mhl$.  The dotted lines indicate the value of
$\tan\beta$ if one moves away from the IFP limit of $Y_t=Y_f$
[see \eq{IFP}]; the upper [lower] dotted line corresponds to
$Y_t/Y_f=0.8$ [$0.9$].
%Predicted value of $\tan\beta$ (dashed line) in the
%IFP scenario as a function of $x\equiv M/m$ with $M_S=1$~TeV
%and $M_t=175$~GeV. The
%solid line depicts the stop mixing parameter $X_t/M_S$ relevant for the
%determination of $\mhl$.
}
\end{figure}
%%%%%%%%%%%%%%%%%%%%%%%%figure%%%%%%%%%%%%%%%%%%%%%%%%

Figure~3 also shows the value of $X_t/M_S$ as a function of $x$.
Recall that both $\tan
\beta$ and $X_t/M_S$ have a crucial impact on $\mhl$.
In particular (assuming that $\tanb\geq 1$ and $|X_t/M_S|\leq
\sqrt{6}$, which is always true in the IFP scenario considered here),
$\mhl$ is an increasing function of both
$\tan\beta$ and $|X_t/M_S|$.  However, as seen from fig.~3,
$\tan\beta$ and $|X_t/M_S|$ do not attain their maximum values at the
same value of $x$, which leads to an effective lowering of the maximum
possible value of $\mhl$.
Moreover, $X_t/M_S$ never reaches the
``maximal value'' of $\left| X_t/M_S\right|=\sqrt{6}$.
This again limits the maximal value of $\mhl$ to lie below its MSSM
upper bound.

The behaviour of $X_t/M_S$ shown in
fig.~3 can be understood by using \eqns{Xt}{Ms}\ plus
the expressions for $\mu$, $A_t$ and the third-generation
scalar squared-masses
given in Section~2 and the appendix.  In the limit where
$M_S\gg m_t$, we obtain
%the approximate expression
\be
\label{XtIFP}
\frac{X_t}{M_S}\simeq\frac{-1.2\,x + \cot\beta (\tan^2\beta-1)^{-1/2}
\left[(1+0.5\tan^2\beta)+(0.5+2\tan^2\beta)\,x^2\right]^{1/2}}
{(0.25+2.75\,x^2)^{1/2}}\,.
\ee
For a typical value of $\tan\beta$ (\eg$\,$ $\tan\beta\sim 1.5$ according
to fig.~3), \eq{XtIFP} reaches a maximum
at $x\sim -0.2$ and lies below the ``maximal value'' $\left|
X_t/M_S\right|=\sqrt{6}$.

%%%%%%%%%%%%%%%%%%%%%%%%figure%%%%%%%%%%%%%%%%%%%%%%%%
\begin{figure}[hbt]
%%\psdraft
\centerline{
\psfig{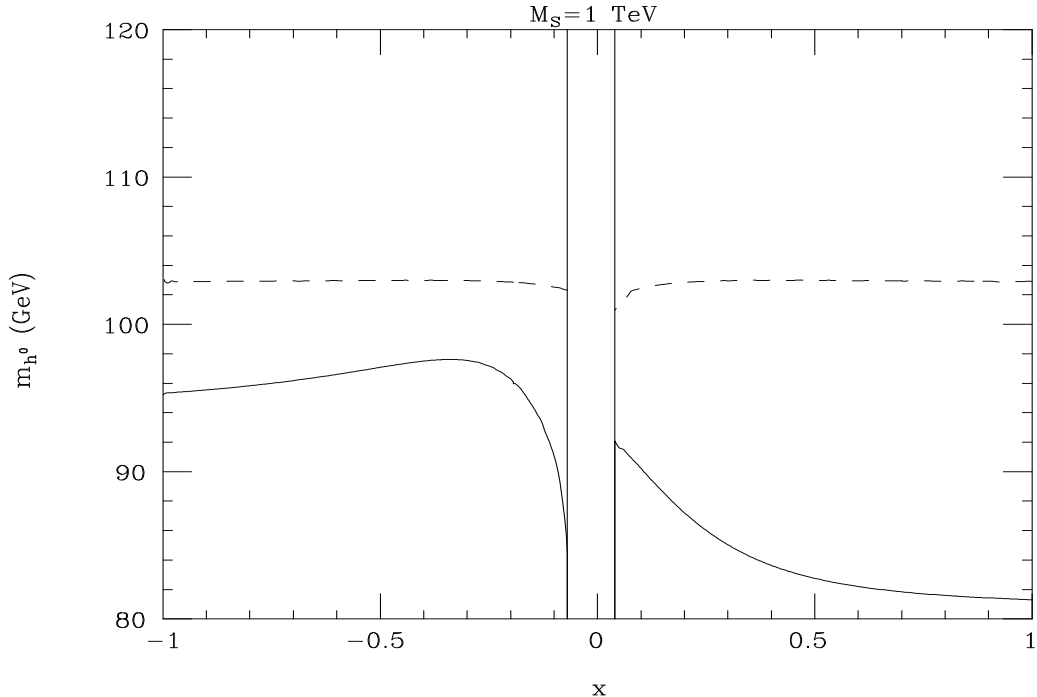}}
\caption{\footnotesize Different approximations to the
lightest CP-even Higgs mass as a function of $x\equiv M/m$ for the IFP
scenario with $M_S=1$~TeV and $M_t=175$~GeV.
The result of the calculation of this paper
is given by the solid curve.  If one omits the stop/gluino loop
corrections to $M_t/m_t(M_t)$ and assumes maximal stop mixing,
one obtains the dashed curved. }
\end{figure}
%%%%%%%%%%%%%%%%%%%%%%%%figure%%%%%%%%%%%%%%%%%%%%%%%%

In fig.~4, the solid curve depicts the results for $\mhl$ in the
IFP scenario considered in this paper, with $M_S=1$~TeV and $M_t=175$~GeV.
Note that the absolute upper bound on $\mhl$ corresponds to
$x\sim -0.3$, although for $x\lsim -0.1$, the variation of
$\mhl$ with $x$ is small.  Numerically the bound on $\mhl$ reads
$\mhl\leq 97$~GeV, with an estimated error of $\pm 2$ GeV
(this error is based on the results of \Ref{hhh}).
In order to illustrate the impact of the new
effects that we have incorporated into the calculation of $\mhl$,
we exhibit the dashed curve in fig.~4.  This latter curve results from a
calculation in which the stop/gluino corrections to the physical (pole)
top-quark mass $M_t$, in terms of the $\overline{\rm DR}$ running mass
$m_t(M_t)$, are omitted [\ie$\,$ taking
$(\Delta m_t/m_t)_{\rm SUSY}=0$], and the stop mixing parameter is set
at its ``maximal" value, $ |X_t/M_S|=\sqrt{6}$.
This procedure has been used
in some works \cite{ceqr,carena,carena2} to deduce an absolute upper
bound on $\mhl$ in the IFP scenario.
In addition, following \Refs{ceqr,carena}, \eq{deltamhs} was used to
obtain the dashed curve for all values of $x$, although we know (see
the discussion near the end of Section~3) that the underlying
assumption of nearly degenerate stops is not appropriate for $|x|\lsim
0.3$.  (The effects of non-degenerate stops were taken into account in
\Ref{carena2}.)
As anticipated, the
dashed curve of fig.~4 significantly overestimates
the $\mhl$ bound over the full $x$ range.
Quantitatively, the overestimate is $\sim 7$~GeV for $x<0$
and $\sim 20$~GeV for $x>0$.

We conclude that previous results for $\mhl$ in the IFP scenario
obtained in the literature had neglected a number of significant
effects, which lead to a substantial reduction in the prediction of the
upper bound for $\mhl$ as a function of the minimal SUGRA parameters.
The Higgs mass upper bounds obtained previously are therefore too
conservative.  The more refined bound of $\mhl\lsim 100$~GeV, obtained
in this paper, is significant in that it lies within the reach of
the LEP-2 Higgs search once the maximum LEP-2 energy and luminosity is
achieved.

%%%%%%%%%%%%%%%%%%%%%%%%figure%%%%%%%%%%%%%%%%%%%%%%%%
\begin{figure}[hbt]
%%\psdraft
\centerline{
\psfig{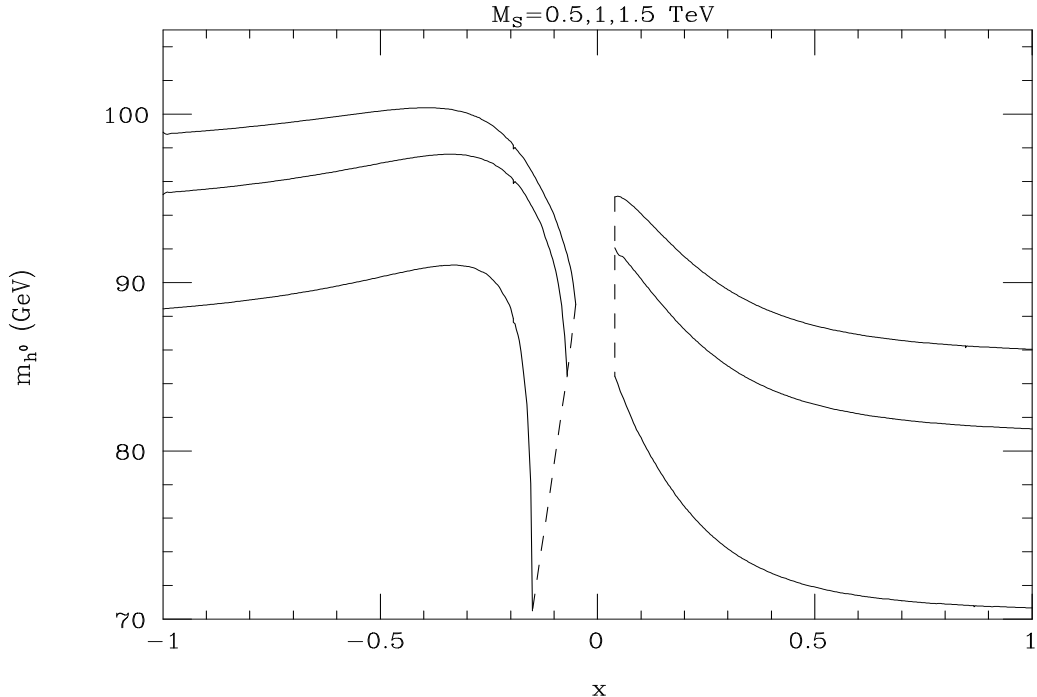}}
\caption{\footnotesize Lightest CP-even Higgs mass as a function
of $x\equiv M/m$ in the IFP scenario for $M_t=175$~GeV and different
values of $M_S$.  The curves shown correspond to 
$M_S=0.5$~TeV (lower curve), $M_S=1$~TeV (middle curve) and $1.5$~TeV
(upper curve). For small $x$, the curves end at the dashed lines where the
stop mass lies below its experimental bound.}
\end{figure}
%%%%%%%%%%%%%%%%%%%%%%%%figure%%%%%%%%%%%%%%%%%%%%%%%%

We next consider the effect of varying the other relevant model
parameters.  Figure~5 shows the value of $\mhl$ as a function of $x$ for
different values of $M_S$; curves for $M_S=$~0.5 TeV, 1 TeV, and 1.5 TeV
are shown.  As expected, the predicted value of $\mhl$ increases
logarithmically with $M_S$. Figure~5 clearly shows a marked asymmetry in
the predicted value of $\mhl$ under a change of sign of $x$.
For $x>0$, the stop mixing contribution to $\mhl$ is less important
since a destructive cancellation takes place between $A_t$ and $\mu$ in
$X_t$
(see fig.~3). As a result, $\mhl$ is typically less than 90~GeV, which
is almost excluded by the current LEP-2 limits on $\mhl$ \cite{sopczak}.
For $x<0$, $A_t$ and $\mu$ have the same sign, thereby enhancing $X_t$.
The corresponding value of $\mhl$ is larger in this case, although for
$M_S\leq 1.5$~TeV, we still predict $\mhl\lsim 100$~GeV.
Larger values for $M_S$ are less plausible, assuming that electroweak
symmetry breaking is a natural consequence of low-energy supersymmetry.

%%%%%%%%%%%%%%%%%%%%%%%%figure%%%%%%%%%%%%%%%%%%%%%%%%
\begin{figure}[hbt]
%%\psdraft
\centerline{
\psfig{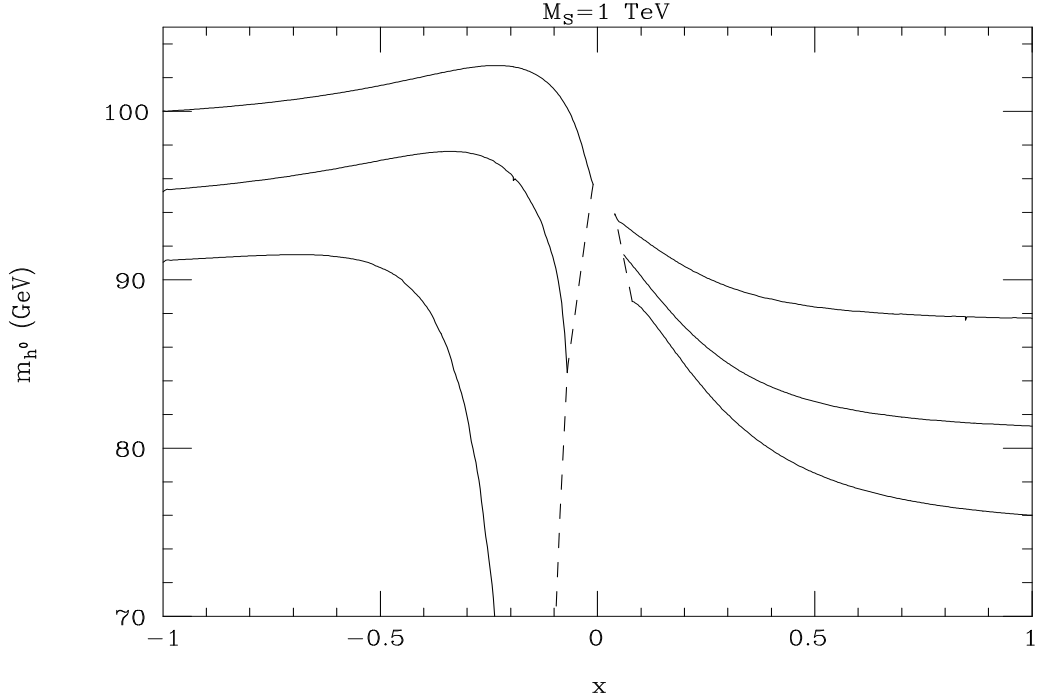}}
\caption{\footnotesize Lightest CP-even Higgs mass as a function
of $x\equiv M/m$ in the IFP scenario for $M_S=1$~TeV and
different values of $M_t$.  The curves shown correspond to
$M_t=170$~GeV (lower curve), $M_t=175$~GeV (middle curve) and $180$~GeV
(upper curve).}
\end{figure}
%%%%%%%%%%%%%%%%%%%%%%%%figure%%%%%%%%%%%%%%%%%%%%%%%%

The predicted value of $\mhl$ is quite sensitive to the value of the
top mass, due to the $m_t^4$ behavior exhibited in \eq{mh1ll}. Taking
into account the experimental error of about 5~GeV in the measured top
quark mass, we exhibit in fig.~6 the predicted value of $\mhl$ as a
function of $x$ for $M_S=1$~TeV and three choices of top quark mass.
For the maximal value of $M_t= 180$~GeV, we see that the predicted upper
bound of $\mhl$ is increased by about 5~GeV, compared to the previous
results shown (for the central value of $M_t=175$~GeV).  We also note
that for $M_t=180$~GeV, the $x$ dependence is somewhat less pronounced,
with the predicted value of $\mhl$ above 88~GeV over the entire allowed
range of $x$. If we impose $M_S\leq 1.5$~TeV,
we conclude that the upper bound of the light CP-even Higgs mass in the
IFP scenario is about 95--105~GeV (for $M_t=175\pm 5$~GeV), although the
upper bound is not saturated  over a significant region of the minimal
SUGRA parameter space.

\section{Conclusions}

The quasi-infrared fixed point model is very attractive, since
a number of quite general and well-motivated initial
conditions lead to a highly predictive low-energy scenario. More
precisely, working within the SUGRA framework (with the
assumption of universality of scalar and gaugino
soft-supersymmetry-breaking masses),
the only two independent parameters for low-energy physics are the
common (high-energy) scalar ($m$) and gaugino ($M$) masses.
%Moreover, the IFP scenario is interesting since it corresponds
%to the low $\tan\beta$ limit of the MSSM.
We have studied
in this framework the value of the light CP-even Higgs mass, $\mhl$,
which is a particularly relevant physical quantity
since it turns out to be greatly constrained.
We have taken into account some important effects that had not
been previously considered. The most notable of these
is the supersymmetric correction to the relation
between the running and the physical top-quark masses, which
lowers the value of $\tan\beta$ and thus that of $\mhl$.
Other effects arise from the precise determination of the stop
mixing parameter $X_t$ (which plays a major role in the
computation of $\mhl$), as well as the observation that
$\tan\beta$ and $X_t$ never conspire to raise $\mhl$ to
its maximum possible value. In addition we have computed $\mhl$ using
the most refined available method, including subdominant
contributions and corrections from stop non-degeneracy.
This substantially reduces the theoretical uncertainty of
our results with respect to previous calculations in the literature.

%%%%%%%%%%%%%%%%%%%%%%%%figure%%%%%%%%%%%%%%%%%%%%%%%%
%\begin{figure}[hbt]
%%\psdraft
%\centerline{
%\psfig{figure=ceh7.ps,height=10cm,width=10cm,angle=90,bbllx=3.cm,%
%bblly=6.cm,bburx=20.cm,bbury=21.cm}}
%\caption{\footnotesize Lightest CP-even Higgs mass as a function
%of $x\equiv M/m$ in an approximate IFP scenario for different values of
%$Y_t/Y_f=1$ (solid curve), $Y_t/Y_f=0.9$ (lower dashed curve) and
%$0.8$ (upper dashed curve).
% $M_S$ is fixed to 1~TeV.}  \end{figure}
%%%%%%%%%%%%%%%%%%%%%%%%figure%%%%%%%%%%%%%%%%%%%%%%%%

Our predictions for $\mhl$ are significantly lower
than previous evaluations, as illustrated in fig.~4.
Figure~5 displays our calculation of $\mhl$ and exhibits its dependence
on $x\simeq M/m$ for different values of
$M_S^2\equiv \half\left( m_{\tilde{t}_1}^2 + m_{\tilde{t}_2}^2\right)$.
For $M_S\leq 1$ TeV and $M_t=175$~GeV we find
$\mhl\le 97\pm 2$~GeV; the upper bound increases slightly for larger
values of $M_S$ and $M_t$.
%Since large values of $M_S$ conflict with
%naturality conditions associated with electroweak symmetry breaking,
For sensible parameter choices, we
conclude that $\mhl\lsim 105$~GeV in the IFP scenario based on the
constrained MSSM with universal scalar and gaugino mass parameters
(as in minimal SUGRA and some superstring models),
and that the most plausible $\mhl$ values may be substantially
smaller. These values of $\mhl$ are within the reach of the LEP-2 Higgs
search.

If LEP-2 fails to discover the $\hl$, then one will be able to
rule out the IFP scenario in the context of SUGRA models with universal
boundary conditions.
Of course, this will not rule out all possible SUGRA models
(or more general versions of the MSSM), where the upper bound
$\mhl\lsim 125$~GeV quoted at the end of Section~3 can still be realized.
Nevertheless, it is worth emphasizing that the IFP scenario does not
correspond merely to a single point of the supersymmetric parameter
space. As noted in Section 1, as long as $Y_t(0)\gsim 0.01$, the
low-energy values of $A_t$ and $\tan\beta$ converge quite closely to
their IFP limits (and independently of the value of
the high-energy parameter $A_0$).  Thus, the IFP prediction of the
Higgs mass bound presented in this paper corresponds to a non-negligible
region of the space of supersymmetric parameters at the high-energy
scale.

Finally, it is interesting to note that the bound on $\mhl$
obtained in this paper is quite robust.  The effect of a small deviation
from the IFP limit leads only to a modest increase in the mass bound of
the light CP-even Higgs boson.  For example, suppose we take
the value of $Y_t$ to lie somewhat below its fixed point value $Y_f$,
but still close enough such that the dependence of $A_t$ on its
high-energy value $A_0$ is negligible.  Then, we find that $\tan\beta$
increases from its predicted IFP value, while $X_t/M_S$ decreases
throughout the
$x$ range of interest.  As a result of these two opposing effects, we
find that the upper bound of $\mhl$ barely shifts (although in contrast
with the results of figs.~4--6, the bound on $\mhl$ as a function of $x$ is
much flatter).  As one takes $Y_t$ further away from the
IFP limit, the dependence of $X_t$ on $A_0$ can no longer be neglected.
One can now attain maximal mixing for reasonable choices of $A_0$.
Since $\tan\beta$ is increased from its IFP value, the upper bound on $\mhl$
also increases.  To illustrate these considerations, we computed the
light CP-even Higgs mass for $M_S=1$~TeV and $m_t=175$~GeV as $Y_t$ is reduced
below $Y_f$.  Using the results for $\tan\beta$ shown by the dotted
curves in fig.~3, we
find that the upper bound on $\mhl$ (which was 97 GeV in the IFP limit)
increases to about 103~GeV [110~GeV], for $Y_t/Y_f=0.9~[0.8]$.   Note that 
away from the IFP limit, the upper
bound on $\mhl$ is nearly independent of the value of $x$ (since the
dependence on $x$ in this case enters mainly through $\tan\beta$).
Thus, if $\hl$ is not discovered at LEP-2,
then one must be somewhat far from the IFP limit examined in this
paper. Given that LEP-2 expects to reach its maximal energy and
luminosity during the next two years, it is safe to say that
the decisive test for the IFP scenario will soon be at hand.

\section*{Acknowledgements}
We are grateful to M. Carena, B. de Carlos, P. Chankowski,
R. Hempfling, S. Pokorski and C.E.M. Wagner for their helpful comments
and suggestions.  We would also particularly like to thank
F. Richard for his probing questions, which led to a more careful
discussion of the ultimate accuracy of the Higgs mass bounds in the IFP
scenario.

\section*{Appendix}

Starting with universal scalar ($m$) and gaugino ($M$) soft masses
at the unification scale
$M_X=[1.2-0.32t_s+0.17t_s^2]\times 10^{16}$~GeV,
\footnote{This result
exhibits the dependence of $M_X$ on the supersymmetric scale $M_S$.
The numerical coefficients of the prefactor are based on a fit
to results obtained from a numerical integration of the two-loop
RGEs, with $\alpha(M_X)=0.039$.}
the soft masses at the supersymmetric scale [of order $M_S$, with
$t_s\equiv \ln(M_S/1~{\rm TeV})$] are:
\be
m_{H_1}^2=m^2+0.5 M^2\nonumber\\
\ee
\be
m_{H_2}^2=m^2+0.5 M^2+\Delta m^2\nonumber\\
\ee
\be
m_{L_i}^2=m^2+0.5 M^2\nonumber\\
\ee
\be
m_{E_i}^2=m^2+0.1 M^2\\
\ee
\be
m_{Q_i}^2=m^2+[4.2-0.69t_s+0.46t_s^2]
M^2+\third\delta_{i3}\Delta m^2\nonumber\\
\ee
\be
m_{U_i}^2=m^2+[3.8-0.69t_s+0.46t_s^2]
M^2+\twothirds\delta_{i3}\Delta m^2\nonumber\\
\ee
\be
m_{D_i}^2=m^2+[3.7-0.69t_s+0.46t_s^2] M^2\,,\nonumber
\ee
where the labels ${H_{1,2}}$ are used for the soft masses of the Higgs
doublets, $L$ for the slepton doublets, $E$ for the singlet sleptons,
$Q$ for the doublet squarks and $U,D$ for up and down singlet squarks
($i$ is a family index), and
\begin{eqnarray}
\Delta m^2=& -\threehalf m^2
{\displaystyle \frac{Y_t}{Y_f}}+
\left([1.6-0.19t_s+0.1t_s^2] A_0 M-\half
A_0^2 \right){\displaystyle \frac{Y_t}{Y_f}}
\left(1-{\displaystyle \frac{Y_t}{Y_f}}\right)\nonumber\\
&+M^2{\displaystyle \frac{Y_t}{Y_f}}
\left([1.3-0.34t_s+0.1t_s^2]{\displaystyle \frac{Y_t}{Y_f}}-
[3.8-0.69t_s+0.6t_s^2]\right),
\end{eqnarray}
where, following ref.~\cite{cw}, we have expressed our results in terms of
the ratio $Y_t/Y_f$.

In addition\footnote{It may seem that \eq{muzero} is in conflict with
\eq{mu} in the quasi-infrared fixed point limit.  Strictly speaking, we
never reach this limit, so $Y_t$ differs from $Y_f$ by a small
amount. Thus, in practice the low-energy value of $\mu$ is first obtained
from \eq{mu}, and then $\mu_0$ is deduced from \eq{muzero}.  Clearly,
$\mu_0\gg\mu$ for $Y_t\simeq Y_f$, which is equivalent to an unnatural
fine-tuning in the electroweak symmetry breaking condition.
Nevertheless, one can still be in the domain of the IFP solution without
significantly violating the naturalness requirements.},
\be \label{muzero}
\mu^2=1.8\mu_0^2\left(1-\frac{Y_t}{Y_f}\right)^{1/2}\,,
\ee
\be \label{bzero}
B=B_0-\half A_0\frac{Y_t}{Y_f}-M\left(0.5-
[0.8-0.1t_s+0.045t_s^2]\frac{Y_t}{Y_f}\right)\,,
\ee
\be
\label{At2}
A_t=A_0\left(1-\frac{Y_t}{Y_f}\right)-M\left(
[2.8-0.31t_s+0.3t_s^2]-[1.6-0.2t_s+0.09t_s^2]\frac{Y_t}{Y_f}\right)
\,.
\ee
In the above, the fitting of the numerical coefficients is accurate in the
range $500~{\rm GeV}\leq M_S \leq 1500$~GeV.
Note that the values of the above parameters at
$M_S$ (particularly those whose running is affected by $\alpha_s$, such
as the squark squared-mass parameters and $A_t$)
are substantially different from the
corresponding values at $\mz$ (see  \Ref{cw}).

\end{document}